\documentclass[aps,showpacs]{revtex4}

\begin{document}

\title{Temperature dependence in random matrix models with pairing
  condensates}

\author{Beno\^\i t Vanderheyden} \affiliation{Department of Electrical
Engineering and Computer Science, \\ Montefiore B{\^a}t.\ B-28,
Universit{\'e} de Li\`ege, \\ B-4000 Li\`ege (Sart-Tilman), Belgium}

\author{A. D. Jackson} \affiliation{ The Niels Bohr Institute,
Blegdamsvej 17, DK-2100 Copenhagen \O, Denmark}

\date{\today}

\begin{abstract}
  
  We address a number of issues raised by a manuscript of Klein,
  Toublan, and Verbaarschot (hep-ph/0405180) in which the authors
  introduce a random matrix model for QCD with two colors, two
  flavors, and fermions in the fundamental representation. Their
  inclusion of temperature terms differs from the approach adopted in
  previous work on this problem (Phys.\ Rev.\ D {\bf 64}, 074016
  (2001).)  We demonstrate that the two approaches are related by a
  transformation that leaves the thermodynamic potential invariant and
  which therefore has no effect on physical observables.

\end{abstract}

\pacs{11.30. Fs, 11.30. Qc, 11.30. Rd, 12.38. Aw}

\maketitle

A number of model calculations of QCD at finite temperature and
non-zero baryon density have revealed the existence of a rich phase
structure~\cite{early-color-sup,recent-color-sup,color-sup-reviews,
HalJacShr98}.  A particularly interesting possibility arises when cold
nuclear matter is brought to densities sufficiently large for quark
degrees of freedom to become important. It is thought that, due to
quark Cooper pairing, cold and dense quark matter behaves like a color
superconductor~\cite{early-color-sup,recent-color-sup}.  The system is
characterized by a variety of color superconducting order parameters
which have been studied as a function of, e.g., the total number of
flavors, the quark masses, the quark chemical potentials, and
temperature~\cite{color-sup-reviews}.  These results have not been
confirmed satisfactorily by lattice simulations. Such calculations are
difficult to perform and interpret at finite baryon densities.  The
fermion determinant appearing in the partition function is complex for
non-zero baryon chemical potential, and standard Monte Carlo methods
can no longer be applied.  As a consequence, our theoretical
understanding of the QCD phase diagram at non-zero baryon density
remains incomplete.

Random matrix theory can provide interesting elementary models that
capture important aspects of the thermodynamic competition among
different order parameters and produce robust results.  Such models
respect the global symmetries of QCD but mimic the detailed dynamics
of the interactions with a form that can be treated analytically.
When supplemented by additional physical insight regarding the form
and relative importance of order parameters, random matrix theory can
efficiently model the physics of systems with non-trivial phase
diagrams. In its usual implementation, the fermion kinetic energy is
ignored; only the interactions and their symmetries are retained.
Such models are thus intrinsically theories of low-energy degrees of
freedom which are treated at the mean field level.  Near critical
regions, random matrix results are similar to those obtained in a
Landau-Ginzburg approach~\cite{HalJacShr98,LG}.

The effects of baryon density and temperature must be included in a
manner that respects the structure of QCD. In the finite temperature 
formalism, the quark chemical potentials, $\mu_f$, and the
temperature, $T$, enter the QCD partition function via the finite
Matsubara frequencies $\mu_f + \omega_n = \mu_f + i (2 n + 1) \pi T$
(with $n=0,\pm 1,\pm 2, \ldots$).  The inclusion of density effects
through the addition of a chemical potential term, $\psi^\dagger \mu
\gamma_4 \psi$, in the Lagrangian is straightforward.  In contrast to
lattice calculations, the resulting random matrix partition function
can be evaluated analytically. Its extremization leads to gap
equations which are polynomial in the condensation fields, and which
can thus be solved analytically or numerically.

Different schemes have been proposed in the literature for the
inclusion of temperature effects.  We note first that, since the
theory describes low-lying energy excitations, much of the critical
physics is captured even if the sum over fermion frequencies is
restricted to the two lowest terms, $\omega_n = \pm i \pi T$.  When
this truncation is adopted in a chiral random matrix theory with
$\mu_f=0$, for example, the resulting model shows a second-order phase
transition independent of the number of flavors~\cite{JacVer96}.
Taking into account all terms in the frequency sum does not change
this result fundamentally.  The sole effect is the introduction of a
temperature mapping that is inhomogeneous (i.e., it depends on $T$)
but monotonic~\cite{JanNowPap98, VanJac01}.  The critical temperature
is thus modified, but neither the topology of the phase diagram nor
the order of the transition are affected.  In the following, we will
only consider cases where the Matsubara sum is truncated to the two
smallest terms.

In an early random matrix model of color superconductivity in QCD with
two flavors~\cite{VanJac00,VanJac01}, we employed a partition function
for which the pairing fields had opposite frequencies,
$\omega_n$. This choice was dictated by the physical observation that
the pairing condensate, $\langle \psi^T (i\sigma_2)_\mathrm{flavor}
(i\sigma_2)_\mathrm{color}(i\sigma_2)_\mathrm{spin}\,\psi\rangle$
(with $i \sigma_2$ the antisymmetric Pauli matrix), is independent of
time.  A Fourier decomposition selects opposite frequencies for the
two flavor fields. This frequency selection is operative independent
of whether one is modeling $SU(3)$-QCD or $SU(2)$-QCD. A seemingly
different temperature dependence, based on the symmetries of the
interactions, was recently introduced by Klein, Toublan, and
Verbaarschot~\cite{KleTouVer03, KleTouVer04}.  The temperature scheme
proposed there reflects the structure of the zero-temperature chiral
random matrix models.  For $SU(2)$ with fundamental fermions, which is
characterized by real matrix elements, the temperature terms are
chosen to be real.  Similarly, both the temperature terms and the
random matrix elements are complex for $SU(3)$ with fermions in the
fundamental representation.

There would thus appear to be two different ways to treat temperature
effects.  The purpose of this paper is to demonstrate that the scheme
based on the symmetries of the interactions is actually equivalent to
that based on frequency selection.  After summarizing the two
temperature schemes in Sec.~\ref{s:T-schemes}, we consider the
transformation that relates them for the case of QCD with two colors
in Sec.~\ref{s:unitary}.  The case of QCD with three colors is
discussed in Sec.~\ref{s:conclusions}.

\section{Temperature schemes}
\label{s:T-schemes}

Random matrix models that are capable of studying the competition
between chiral, diquark, and other condensation channels are
extensions of the chiral random matrix
models~\cite{ShuVer93,VerZah93,JacVer96,HalJacShr98}. In the sector of
zero topological charge and for $N_f$ flavors of fermions with masses
$m_f$, the partition function has the form
\begin{eqnarray}
Z_0 & = & \prod_f \int {\cal D}\psi_f^\dagger {\cal D}\psi_f {\cal D} W \,
e^{- \frac{n \beta \Sigma^2}{2}\,\,\mathrm{Tr}\,W W^\dagger}\,\,
\exp\left( \sum_{f=1}^{N_f} \psi_f^\dagger (D[W] + m_f)
\psi_f^{\phantom \dagger} \right),
\label{Zchiral}
\end{eqnarray}
where $W$ is an $n \times n$ matrix which models the interaction.  The
elements of $W$ are drawn on a Gaussian distribution of inverse
variance $\Sigma$.  Here, $n$ is a measure of the number of low energy
degrees of freedom and is to be taken to infinity at the end of the
calculations (i.e., in the thermodynamic limit). The Dirac operator,
$D[W]$, respects the chiral symmetry of QCD, $\{D,\gamma_5\}=0$. In
the basis of the eigenstates of $\gamma_5$, $D[W]$ can be decomposed
as
\begin{eqnarray}
D[W] = \left(
\begin{array}{cc}
0 & i W\\
i W^\dagger & 0 \\
\end{array}
\right).
\label{Dirac}
\end{eqnarray}

For random matrix models of QCD with $SU(3)$ and fermions in the
fundamental representation, $W$ is complex. This choice corresponds to
a chiral unitary ensemble characterized by a Dyson index $\beta =
2$. The QCD Dirac operator in $SU(2)$ with fermions in the fundamental
representation satisfies an additional anti-unitary symmetry,
\begin{eqnarray}
[C(\sigma_2)_{\mathrm{color}} K,i D] = 0,
\label{antiunitary}
\end{eqnarray}
with $\left(C(\sigma_2)_{\mathrm{color}} K \right)^2=1$.  ($C$ is the
charge conjugation operator, $\sigma_2$ the antisymmetric color
matrix, and $K$ the complex conjugation operator.)  This leads to a
particular basis of states in which $D$ is real. Accordingly, $W$ is
chosen real in chiral random matrix models for $SU(2)$, and the model
belongs to the chiral orthogonal ensemble with an index $\beta =
1$~\cite{VerPRL94}.  The essential difference between the unitary and
orthogonal ensembles lies in the number of independent random
variables that are allowed per matrix element.  (I.e., A complex
number has two degrees of freedom; a real number has only one.)  This
difference is important in determining the statistical properties of
the Dirac operator and, in turn, influences the phase diagram.

The inclusion of quark chemical potentials, $\mu_f$, and the
temperature, $T$, follows from the frequency representation of the
partition function in the finite temperature
formalism~\cite{JacVer96,finiteTformalism}. The path integral is
evaluated for fermion fields that obey antiperiodic boundary
conditions in Euclidean time. Thus, the Fourier decomposition
\begin{eqnarray}
\psi(\tau) = T \sum_{\omega_n} \, e^{- \omega_n \tau} \psi(\omega_n)
\label{Fourier-decomposition}
\end{eqnarray} 
contains only the odd Matsubara frequencies
\begin{eqnarray}
\omega_n = i (2 n + 1) \pi T.
\end{eqnarray} 
The dependence on $\mu_f$ and $T$ appears as a component of the
Lagrangian of the form
\begin{eqnarray}
L_1 = \int d^4 x \sum_{f}\, \psi^\dagger_f \,\gamma_4 (\partial_\tau +
\mu_f ) \psi_f,
\end{eqnarray}
whose Fourier representation is given as
\begin{eqnarray}
L_1 = T \sum_n \int d^3 \vec{x}\,\sum_f \psi^\dagger_{f}(\omega_n)
\,\gamma_4 (- i (2 n + 1) \pi T + \mu_f)
\psi_f^{\phantom{\dagger}}(\omega_n).
\label{mufT-QCD}
\end{eqnarray}
There is a certain freedom associated with the introduction of such a
term in random matrix models. Equation~(\ref{mufT-QCD}) suggests that
we should add a deterministic term with the block structure of
$\gamma_4$ to the Dirac operator of Eq.~(\ref{Dirac}).  With Euclidean
matrices, this gives
\begin{eqnarray}
D[W] \to D[W;\mu_f,T]  = \left(
\begin{array}{cc}
0 & i W + \omega_+(\mu_f,T)\\ i W^\dagger + \omega_-(\mu_f,T) & 0 \\
\end{array}
\right),
\label{Dirac-muT}
\end{eqnarray}
where $\omega_\pm$ are $n \times n$ matrices. Here, $n$ is a compound
index which combines random matrix and Matsubara frequency indices.
Because the ensemble of matrices $W$ is invariant under unitary
($\beta = 2$) or orthogonal ($\beta = 1$) basis transformations, many
equivalent forms of $\omega_\pm$ can be adopted, and their detailed
form cannot be specified uniquely. (We will also consider another
transformation below.)  We know, however, that the eigenvalues of
$\omega_\pm$ must be those of the operator in Eq.~(\ref{mufT-QCD}). In
the following, we restrict ourselves to the two lowest Matsubara
frequencies, $\omega_n=\pm i \pi T$. The appropriate forms of
$\omega_\pm$ must then have two series of $n/2$-degenerate eigenvalues
$\mu_f \pm i \pi T$.

We now turn to the various schemes proposed for implementing
temperature dependence.  In an early model of diquark
condensation~\cite{VanJac00, VanJac01} in QCD with two flavors, we
considered random matrix interactions that included spin and color
quantum numbers explicitly.  This choice leads to an expanded block
structure
\begin{eqnarray}
W = \sum_{\mu=0}^3 \sum_{a=1}^3 \left(\sigma_\mu \otimes
\sigma_a\right) A_{\mu a},
\label{W-exploded}
\end{eqnarray}
where $\sigma_\mu = (1, i \vec\sigma)$ represents spin degrees of
freedom with the Pauli matrices $\vec{\sigma}$, while $\sigma_a$ are
the $SU(2)$ color matrices. The interactions are to be integrated over
the real matrices $A_{\mu a}$ whose elements are drawn on a Gaussian
distribution.  The $W$ appearing in Eq.~(\ref{W-exploded}) is complex
independent of the number of colors.  Because the dependence on spin
and color is deterministic, the model nevertheless has the appropriate
number of degrees of freedom per matrix element.  For example, for
$N_c = 2$ the Dirac operator $D[W]$ in Eq.~(\ref{Dirac}) with the
block structure of Eq.~(\ref{W-exploded}) satisfies the antiunitary
symmetry of Eq.~(\ref{antiunitary}). More generally, we have shown
that for $N_c=3$ and $N_c=2$ the eigenvalues of the Dirac operator
$D[W]$ exhibit the spectral correlations expected for the chiral
unitary and orthogonal ensembles, respectively~\cite{VanJac03}.

The deterministic part of the Dirac operator is given as
\begin{eqnarray}
\omega_+ (\mu_f,T) = \omega_-(\mu_f,T) = & ({\bf 1}_\mathrm{spin} \otimes
{\bf 1}_\mathrm{color}) \left( \mu_f + i
s_f \pi T \right),
\label{our-T-dependence}
\end{eqnarray}
where the choice of $s_f=\pm 1$ follows a selection rule. First, $s_f$
is taken positive for flavor $f=1$ and negative for flavor $f=2$. The
contribution from the opposite choice, $s_1 = -1$ and $s_2 = 1$, is
then added to obtain the thermodynamic potential.  (In
Refs.~\cite{VanJac00, VanJac01}, we considered the limit $\mu_1 =
\mu_2$; in that case, the two contributions are identical and the sum over
pairs amounts to a factor of two.)  This procedure is suggested by the
Fourier expansion of the condensates in a microscopic theory. The
chiral condensate is given by the equal time correlator
$C_{\sigma}\sim \langle \bar\psi(\tau) \psi(\tau)\rangle$, which is
also time independent. A frequency decomposition of the fields,
Eq.~(\ref{Fourier-decomposition}), leads to an expression which
contains a sum over all frequency pairs. Since the correlator
$C_\sigma$ is time independent, the only non-vanishing terms are those
for which the Fourier components ${\bar{\psi}}(\omega_n)$ and
$\psi(\omega_m)$ have equal frequencies (or $m=n$): $C_{\sigma}\sim T
\langle \sum_{n} {\bar\psi}(\omega_n) \psi(\omega_n) \rangle$.
Similarly, the diquark condensate is given by the equal-time
correlator $C_{\Delta} \sim \langle \psi^T(\tau) (i
\sigma_2)_{\mathrm{color}}(i\sigma_2)_{\mathrm{spin}}(i
\sigma_2)_{\mathrm{flavor}}\,\psi(\tau)\rangle$ and is constant in
time $\tau$. A Fourier decomposition now selects opposite frequencies
for the two fields, which necessarily correspond to different flavors
because of $(i \sigma_2)_{\mathrm{flavor}}$. This mechanism was
implemented in the random matrix model through the selection of
appropriate pairs $(s_1,s_2)$.

In recent work~\cite{KleTouVer03, KleTouVer04}, Klein, Toublan, and
Verbaarschot treated the chiral random matrix ensembles, studying
diquark condensation for the case $N_c = 2$ only. The zero temperature
and zero $\mu_f$ version of their model follows Eq.~(\ref{Zchiral})
with $W$ complex for the unitary ensemble and real for the orthogonal
ensemble.  In their approach, the dependence on $\mu_f$ and $T$
follows the symmetries of the chiral ensembles. For the unitary
ensembles, $\omega_\pm$ are complex. In the appropriate sub-basis of
frequency states, they assume the diagonal form
\begin{eqnarray}
\omega_+(\mu_f,T) = \omega_-(\mu_f,T) = \left(\begin{array}{cc} \mu_f
 + i\pi T & 0 \\ 0 & \mu_f - i \pi T \end{array}\right),
\label{T-omega3}
\end{eqnarray}
where each block matrix has dimension $n/2 \times n/2$. For the chiral
orthogonal ensembles, the antiunitary symmetry of
Eq.~(\ref{antiunitary}) is imposed, which leads to a particular basis
for which the $\omega_\pm$ matrices are real. Ensuring that their
eigenvalues are given as $\mu_f \pm i \pi T$, the proposed dependence
is 
\begin{eqnarray}
\omega_+ & = & \left(\begin{array}{cc} \mu_f & \pi T \\ - \pi T & \mu_f
\end{array}\right), \nonumber \\ \omega_- & = & \omega_+^T,
\label{T-omega2}
\end{eqnarray} 
where the dimensions of the block matrices are again $n/2 \times n/2$.

At first sight, the two forms of $\omega_\pm$, Eqs.~(\ref{T-omega3})
and (\ref{T-omega2}), might seem non-equivalent and also different
from that of Eq.~(\ref{our-T-dependence}).  There thus appears to be a
disagreement between the temperature dependence that is obtained from
a generic structure suggested by symmetry and one which is based on
physical knowledge of the frequency couplings that are allowed. In the
next section, we consider the case $N_c = 2$ and show that the
dependence expressed by the selection rule in
Eq.~(\ref{our-T-dependence}) leads to the same effective thermodynamic
potential as that obtained with the dependence of
Eq.~(\ref{T-omega2}).  We comment on the correct form to be used for
$N_c = 3$ in Sec.~\ref{s:conclusions}.

\section{Equivalences between the temperature schemes}

\label{s:unitary}

It is useful to compare the temperature schemes of
Eqs.~(\ref{our-T-dependence}), (\ref{T-omega3}), and (\ref{T-omega2}) 
in a common framework. To this end, we recast the model developed in
\cite{VanJac00,VanJac01} in a form close to that of
Refs.~\cite{KleTouVer03,KleTouVer04}. We consider a model for $N_c=2$,
two flavors, and zero quark masses. We work with an interaction $W$
having the expanded block-structure of Eq.~(\ref{W-exploded}) and
derive the thermodynamic potential for the partition function of
Eq.~(\ref{Zchiral}). After integrating over the random matrices
$A_{\mu a}$ and performing a Fierz transformation, we obtain the
partition function~\cite{VanJac00}
\begin{eqnarray}
Z = \prod_f \int {\cal D}\psi_f {\cal D} \psi^\dagger_f \exp\left(Y\right)
\exp\left(\sum_{f=1,2}\psi^\dagger_f 
\left(
\begin{array}{cc}
0 & \omega_+(\mu_f,T) \\
\omega_-(\mu_f,T) & 0 \\
\end{array}
\right)
 \psi_f\right),
\label{Z}
\end{eqnarray}
where $Y$ represents the four-point interaction. If only the chiral and
diquark channels are retained,
\begin{eqnarray}
Y & = & \frac{3}{4 n \beta \Sigma^2}\, \left(
   \left(\psi^\dagger_{R1}\cdot \psi^{\phantom{\dagger}}_{R1}\right)
   \left(\psi^\dagger_{L1}\cdot \psi^{\phantom{\dagger}}_{L1}\right) + 
   \left(\psi^\dagger_{R2}\cdot \psi^{\phantom{\dagger}}_{R2}\right)
   \left(\psi^\dagger_{L2}\cdot \psi^{\phantom{\dagger}}_{L2}\right) + 
\right. \nonumber \\
    && \left.  \left(\psi^\dagger_{R1}\cdot P_\Delta \cdot \psi_{R2}^*\right)
   \left(\psi^T_{L2}\cdot P_\Delta \cdot \psi^{\phantom{\dagger}}_{L1}\right) +
   \left(\psi^\dagger_{L1}\cdot P_\Delta \cdot \psi_{L2}^*\right)
   \left(\psi^T_{R2}\cdot P_\Delta \cdot \psi^{\phantom{\dagger}}_{R1}\right)
\right).
\label{RR-LL}
\end{eqnarray}
Here, we have adopted the notation $\phi\cdot\chi \equiv \sum_{i=1}^n
\phi_i \chi_i$, where $i$ is the random matrix index, and have
introduced the projector $P_\Delta = (i\sigma_2)_\mathrm{spin}
(i\sigma_2)_\mathrm{color}$.

Following \cite{JacVer96} and \cite{KleTouVer04}, bosonic auxiliary
variables are introduced via a Hubbard-Stratonovich transformation
that accounts for the different combinations of spin, color, and
flavor quantum numbers. This transformation is given by
\begin{eqnarray}
e^Y = \int {\cal D}A \, e^{- {n \beta \Sigma^2\over 6}\,\, 
{\mathrm{Tr}}\, A A^\dagger}\,e^{{\cal Y}(A)},
\label{Hubbard}
\end{eqnarray}
where $A$ is a complex, antisymmetric $16 \times 16$ matrix, and ${\cal Y}(A)$
is the bilinear
\begin{eqnarray}
{\cal Y}(A) &=& \sum_{f=1,2} \, {1\over 2}\,
\left(\begin{array}{c} \psi^f_R \\ \psi^{f*}_R \\ \end{array}\right)^T
(A^\dagger \otimes {\mathbf{1}}_n )
\left(\begin{array}{c} \psi^f_R \\ \psi^{f*}_R \\ \end{array}\right) 
+ \sum_{f=1,2} \, {1 \over 2}\,
\left(\begin{array}{c} \psi^f_L \\ \psi^{f*}_L \\
\end{array}\right)^T
( 
B\otimes \mathbf{1}_n
)
\left(\begin{array}{c} \psi^f_L \\ \psi^{f*}_L \\ 
\end{array}\right).
\label{YA}
\end{eqnarray}
Here, ${\bf 1}_n$ represents the identity matrix in the combined space
of random matrix/frequency indices and $B$ is given as
\begin{eqnarray}
B = \left(\begin{array}{cc} 0 & 1 \\ 1 & 0 \end{array} \right) A
\left(\begin{array}{cc} 0 & 1 \\ 1 & 0 \end{array} \right).
\end{eqnarray}
In a basis of states $(\psi_f^{\phantom{*}}, \psi_f^*) = (\psi_1^{\phantom{*}},
\psi_2^{\phantom{*}}, \psi_1^*, \psi_2^*)$, the matrix $A$ assumes the
form
\begin{eqnarray}
A=\left(\begin{array}{cccc} 0 & \Delta_R P_\Delta & \sigma_1^* & 0 \\
-  \Delta_R P_\Delta & 0 & 0 & \sigma_2^* \\
- \sigma_1^* & 0 & 0 & -  \Delta_L^* P_\Delta \\
0 & - \sigma_2^* &   \Delta_L^* P_\Delta & 0\end{array}\right),
\label{A}
\end{eqnarray}
where each block is $4 \times 4$ (two spins $\times$ two colors), and
the auxiliary variables $\Delta_L, \Delta_R, \sigma_1,$ and $\sigma_2$
are complex scalars.

The temperature and chemical potential terms of Eq.~(\ref{Z}) can be
arranged as
\begin{eqnarray}
\sum_{f=1,2}\psi^\dagger_f 
\left(
\begin{array}{cc}
0 & \omega_+ \\
\omega_- & 0 \\
\end{array}
\right)
\psi_f^{\phantom{\dagger}} & = & \sum_{f=1,2} \, {1 \over 2}
\,
\left(\begin{array}{c} \psi^f_R \\ \psi^{f*}_R \\ 
\end{array}\right)^T 
\left(\begin{array}{cc}0 & - \omega_+^T \\ 
\omega_+ & 0 \\ \end{array}\right)
\left(\begin{array}{c} \psi^f_L \\ \psi^{f*}_L \\ 
\end{array}\right) \nonumber \\
&& + \sum_{f=1,2}\, {1 \over 2}
\,
\left(\begin{array}{c} \psi^f_L \\ \psi^{f*}_L \\ 
\end{array}\right)^T 
\left(\begin{array}{cc}0 & - \omega_-^T \\ 
\omega_- & 0 \\ \end{array}\right)
\left(\begin{array}{c} \psi^f_R \\ \psi^{f*}_R \\ 
\end{array}\right).
\label{LR-RL}
\end{eqnarray}

Combining Eqs.~(\ref{Z}), (\ref{Hubbard}), and (\ref{LR-RL}) and
integrating over the fermion fields, we finally obtain
\begin{eqnarray}
Z & = &\int {\cal D} A \, e^{-{n \beta \Sigma^2 \over 6}\,\,\mathrm{Tr}\,A
 A^\dagger} 
\exp{\left({1\over 2}\,\mathrm{Tr} \log \left(
 \begin{array}{cc}
   A^\dagger \otimes {\bf 1}_n & 
               \left(\begin{array}{cc} 0 & -\omega_+^T \\
                            \omega_+ & 0 \\ 
                     \end{array}\right) \\
 \left(\begin{array}{cc} 0 & - \omega_-^T \\
                            \omega_- & 0 \\ 
                     \end{array}\right)
 & 
B \otimes \mathbf{1}_n \\
 \end{array}
\right) \right)}.
\label{ZA}
\end{eqnarray}
For later discussions, it is important to note that the diagonal
blocks of the argument of the logarithm are given as an exterior
product with ${\bf 1}_n$. Different Matsubara frequencies will thus
correspond to the same auxiliary matrix, $A^\dagger$ or $B$.

Equation~(\ref{ZA}) is the central expression which we will use to
compare the different temperature schemes. We will concentrate on the
same condensation channels as those considered earlier~\cite{VanJac00,
VanJac01}.  We take equal quark chemical potentials, $\mu_1 = \mu_2 =
\mu$.  The chiral condensate is considered to be flavor independent so 
that $\sigma_1 = \sigma_2 = \sigma$. Moreover, we have $\langle
\psi_{2R}^T P_\Delta \psi_{1R}^{\phantom{T}} \rangle = \langle
\psi_{2L}^T P_\Delta \psi_{1L}^{\phantom T}\rangle$, and the auxiliary
fields satisfy the equality $\Delta_R=\Delta_L=\Delta$.

\subsection{Selection rule scheme}

The temperature dependence proposed in Refs.~\cite{VanJac00,VanJac01}
can now be represented as ($\mu_1 = \mu_2 = \mu$)
\begin{eqnarray}
\left\{
\begin{array}{lclcl}
\omega_+(\mu_1,T) & = & \omega_-(\mu_1,T) & = &
\left(\begin{array}{cc}\mu + i \pi T & 0 \\ 0 & \mu -i \pi T \\
\end{array}\right)_\mathrm{Matsubara} 
\\
\omega_+(\mu_2,T) & = & \omega_-(\mu_2,T) & = &
\left(\begin{array}{cc}\mu - i \pi T & 0 \\ 0 & \mu + i \pi T \\
\end{array}\right)_\mathrm{Matsubara}. \\
\end{array}
\right.
\label{T-case1}
\end{eqnarray}
Inserting this form in Eq.~(\ref{ZA}) leaves us with a straightforward
evaluation of a trace.  We obtain
\begin{eqnarray}
Z & = & \int {\cal D} A \,\,e^{-4 n \Omega},
\end{eqnarray}
with an effective thermodynamic potential 
\begin{eqnarray}
\Omega(\sigma,\Delta) & = & {\frac{2 \beta \Sigma^2}{3}}
 \left(\sigma^2 + |\Delta|^2\right) - \log\left((\sigma - \mu)^2 +
 |\Delta|^2 + \pi^2 T^2 \right) - \log\left((\sigma + \mu)^2 + |\Delta|^2 +
 \pi^2 T^2\right).
\label{Omega}
\end{eqnarray}
(The inclusion of a quark mass $m_f=m$ would amount to replace
$\sigma$ by $\sigma + m$ in the argument of the logarithm.)  In the
thermodynamic limit $n\to \infty$, one can evaluate the partition
function integral exactly by a saddle point method and obtain
\begin{eqnarray}
\lim_{n\to \infty} Z \sim e^{-4 n
\Omega(\sigma_{\mathrm{saddle}},\Delta_\mathrm{saddle})}
\end{eqnarray}
where the fields $\sigma_{\mathrm{saddle}}$ and
$\Delta_{\mathrm{saddle}}$ are solutions of the gap equations
\begin{eqnarray}
\frac{\partial \Omega}{\partial \sigma}(\sigma,\Delta) & = & 0,\\
\frac{\partial \Omega}{\partial \Delta}(\sigma,\Delta) & = & 0.
\end{eqnarray}
Although the potential in Eq.~(\ref{Omega}) has been obtained using a
formalism different from that used in Ref.~\cite{VanJac01}, we recover
the earlier result. We now turn to other temperature schemes.

\subsection{A selection rule scheme with explicit flavor symmetry}
\label{s:srule}

At first sight, the temperature dependence in Eq.~(\ref{T-case1})
might seem to violate flavor symmetry. However, this is not the
case. The sign reversal in $T$ for flavor $2$ can also be expressed as
a permutation of the two Matsubara frequencies, since
\begin{eqnarray}
- \left(\begin{array}{cc} i \pi T  & 0 \\ 0 & -i \pi T \\\end{array}\right) = 
\left(\begin{array}{cc} 0 & 1 \\ 1 & 0 \\\end{array}\right) 
\left(\begin{array}{cc} i \pi T & 0 \\ 0 & -i \pi T \\\end{array}\right)
\left(\begin{array}{cc} 0 & 1 \\ 1 & 0 \\\end{array}\right). 
\end{eqnarray}
Therefore, we can go to an equivalent representation by permuting the
frequency states for flavor $2$. The selection of opposite frequencies
applies to diquark and antidiquark bilinears for which the projector
$P_\Delta$ is sandwiched between the flavor fields as in
Eq.~(\ref{RR-LL}). Hence, the effect of permuting positive and
negative frequencies for flavor $2$ transforms $P_\Delta$ into
\begin{eqnarray}
P_\Delta \to \tilde{P}_\Delta = P_\Delta \left(
\begin{array}{cc} 0 & 1 \\ 1 & 0 \\ \end{array}\right)_\mathrm{Matsubara}
= (i\sigma_2)_\mathrm{spin}\otimes (i
\sigma_2)_\mathrm{color} \otimes \left(
\begin{array}{cc} 0 & 1 \\ 1 & 0 \\ \end{array}\right)_\mathrm{Matsubara}.
\label{srule}
\end{eqnarray}
By contrast, the chiral bilinears are unaffected by the frequency
permutation, e.g., $\psi^{2\dagger}\psi^2 = \psi^{2\dagger}_+ \psi^2_+
+ \psi^{2\dagger}_- \psi^2_- $.  The sole effect of permuting the
frequencies for flavor $2$ is thus to make the frequency selection
mechanism appear in the structure of the order parameter through the
projector $\tilde{P}_\Delta$. The frequency component of this operator
is flavor diagonal and $\tilde{P}_\Delta$ therefore respects the required
flavor symmetries.

\subsection{Antisymmetric temperature dependence}

The apparent disagreement between a temperature dependence based on
symmetry and one based on selecting frequency couplings would
immediately be resolved by a transformation relating
Eqs.~(\ref{T-omega2}) and~(\ref{our-T-dependence}).  In the
antisymmetric representation of Eq.~(\ref{T-omega2}), the basis of
states is meant to be that for which all matrix elements of $W$ are
real. This is certainly not the basis set to be used in our model,
Eq.~(\ref{Z}), since the spin and color dependence of
Eq.~(\ref{W-exploded}) makes $W$ complex. Nevertheless, the Dirac
operator $D[W]$ of Eq.~(\ref{Dirac}) satisfies the anti-unitary
symmetry of Eq.~(\ref{antiunitary}), and we could in principle choose
the appropriate spin and color basis to obtain a real
$W$~\cite{VanJac01}.  This is a consequence of the fact that the
matrix elements contain the proper number of random degrees of
freedom. In the following we are interested in the transformation
properties of the deterministic temperature terms $\omega_\pm$ and
thus do not consider transformations in spin/color space. We focus
instead on the effects of a partial transformation acting only on
frequency indices.

We wish to establish the relationship between the temperature
dependence based on the selection rule of Eq.~(\ref{srule}) and an
antisymmetric dependence. We start from the representation of
Sec.~\ref{s:srule}, for which
\begin{eqnarray}
\omega_\pm
= \mu + \left(\begin{array}{cc} i \pi T & 0 \\ 0 & - i \pi
T \end{array}\right),
\end{eqnarray}
and the diquark order parameter contains the projector $\tilde{P}_\Delta$ of
Eq.~(\ref{srule}).  We rewrite the diagonal temperature form as 
\begin{eqnarray}
\left(
  \begin{array}{cc} i \pi T & 0 \\ 0 & -i \pi T \\ \end{array}
\right)
=
U_m 
\left(
  \begin{array}{cc} 0 & \pi T \\ -\pi T & 0\\ \end{array}
\right) U_m^\dagger,  
\end{eqnarray}
where
\begin{eqnarray}
U_m = {1 \over \sqrt{2}}\, 
\left(
\begin{array}{cc} 1 & -i \\ 1 &  i \end{array}
\right)
\end{eqnarray}
is unitary.  We then perform the transformation 
\begin{eqnarray}
\psi_{R,L} & = & U \, \psi_{R,L}^{'}, \\
\psi^*_{R,L} & = & U^* \, \psi^{'*}_{R,L}, 
\end{eqnarray}
with 
\begin{eqnarray}
U = \left(
       \begin{array}{cc}
        U_m \otimes \mathbf{1}_{n/2} & 0 \\ 0 & U_m \otimes \mathbf{1}_{n/2}
       \end{array}
    \right)
\otimes {\bf 1}_{\mathrm{spin}} \otimes {\bf 1}_{\mathrm{color}}
\otimes {\bf 1}_{\mathrm{flavor}}.
\label{unitary-transformation}
\end{eqnarray}
This transformation leaves the partition function of Eq.~(\ref{ZA})
unaffected.  Since $U$ only acts in frequency space, it does not affect
the quadratic term, $\mathrm{Tr}\,A A ^\dagger$.  The logarithmic term
is also unaltered since $\mathrm{Tr}\log M = \log \mathrm{det} M$, and
the determinant is invariant under the transformation
\begin{eqnarray}
M \to M^{'} = 
\left(\begin{array}{cc} U^T & 0 \\ 0 & U^\dagger \\ \end{array}\right) 
M 
\left(\begin{array}{cc} U & 0 \\ 0 & U^* \\ \end{array} \right).
\end{eqnarray}
Hence, the transformation, $U$, brings the temperature representation
of Sec.~\ref{s:srule} to an equivalent representation with an
antisymmetric temperature dependence.

We now examine the transformation of the bilinear form of
Eq.~(\ref{YA}) bearing in mind that we started with the
representation of Sec.~\ref{s:srule}, for which the diquark terms
contain the projector $\tilde{P}_\Delta$ of Eq.~(\ref{srule}). 
Mesonic bilinears of the form
$\psi^\dagger_i \psi_j$ (where $i$ and $j$ represent some combination 
of spin/color/flavor indices) are invariant under the $U$-transformation as 
\begin{eqnarray}
\psi^{\dagger}_i \psi_j = \psi^{\dagger'}_i U^\dagger U \psi_j^{'} =
\psi^{\dagger'}_i \psi_j^{'}.
\label{mesonic}
\end{eqnarray}
By contrast, diquark bilinears of the form $\psi^T_i \tilde{P}_\Delta
\psi_j$ are transformed to
\begin{eqnarray}
\psi^T_i \tilde{P}_\Delta \psi_j =  \psi^{'T}_i P^{'}_\Delta \psi_j^{'},
\end{eqnarray}
with
\begin{eqnarray}
P^{'}_\Delta = U^T_m  \tilde{P}_\Delta  U_m =P_\Delta,
\end{eqnarray}
where again, $P_\Delta = (i\sigma_2)_\mathrm{spin}
(i\sigma_2)_\mathrm{color}$ is block diagonal in frequency space.

To summarize, the temperature dependence in the new basis is
antisymmetric, while the diquark order parameter no longer contains an
explicit frequency selection operator. This is precisely the form
proposed in Ref.~\cite{KleTouVer04}.  We thus have shown that it is
equivalent to the selection rule of Eq.~(\ref{srule}) and, by
extension, to the one originally proposed in
Ref.~\cite{VanJac00,VanJac01}.  To complete the proof, it is readily 
verified that the various temperature schemes of
Eqs.~(\ref{our-T-dependence}), (\ref{T-omega2}), and (\ref{srule}),
used with the corresponding forms for the diquark projector
($P_\Delta$, $\tilde{P}_\Delta$, or $P_\Delta^{'}=P_\Delta$), all
yield the thermodynamic potential of Eq.~(\ref{Omega}).

\subsection{Pion condensates}

Although the temperature dependence in Eq.~(\ref{our-T-dependence})
produces the correct result for modelling diquark and chiral
condensation, it is inconvenient if one also wishes to consider the
possibility of forming a pion condensate, $\langle \psi^\dagger
\gamma_5 (i\sigma_2) \psi\rangle$. Pion bilinears must be taken
between fields of equal frequencies and opposite flavors, whereas
elements associated with diquark bilinears require opposite
frequencies and opposite flavors. Considering both channels 
thus leads to some tedious bookkeeping. However, the equivalent
form of Eq.~(\ref{srule}) can be implemented much more easily and
extended to the inclusion of pion bilinears. The resulting form takes
a diagonal temperature dependence, considers diquark bilinears with an
off-diagonal projector $\tilde{P}_\Delta$, while it leaves both the
chiral and pion bilinears diagonal in frequency space.  We now show
that such prescription gives the same results as those obtained with
an antisymmetric temperature dependence.

A Fierz projection of the random matrix interactions on the pion
condensation channel yields a four fermion component
\begin{eqnarray}
Y_\pi & = & \frac{3}{32 n \beta \Sigma^2}\,(\psi^\dagger \gamma_5 
(i\sigma_2)_{\mathrm{flavor}}
\psi)^2.
\end{eqnarray}
Using again a Hubbard-Stratonovich transformation to introduce an
associated auxiliary variable, $\rho$, one is led to consider the
auxiliary matrix
\begin{eqnarray}
A=\left(\begin{array}{cccc} 0 & \Delta_R P_\Delta & 
\sigma_1^* & \rho \\
-  \Delta_R P_\Delta & 0 & - \rho & \sigma_2^* \\
- \sigma_1^* & \rho & 0 & -  \Delta_L^* P_\Delta \\
- \rho & - \sigma_2^* &   \Delta_L^* P_\Delta & 0\end{array}\right).
\label{Arho}
\end{eqnarray}
As previously, the frequency selection rule is introduced by replacing
$P_\Delta$ by $\tilde{P}_\Delta$ in the diquark and antidiquark bilinears
of Eq.~(\ref{YA}) or equivalently in the argument of the logarithm of
Eq.~(\ref{ZA}). Such a temperature scheme is completely consistent 
with the spirit with which we derived the temperature dependence in
Refs.~\cite{VanJac00,VanJac01}; we have verified that it produces the same
thermodynamical potential as that obtained with an antisymmetric
temperature dependence.  The result is identical to that of
Ref.~\cite{KleTouVer04}.

\section{Discussion and conclusions}
\label{s:conclusions}

For $N_c=2$, we have resolved the apparent disagreement between the
dependence based on an antisymmetric temperature term and that based
on a frequency selection rule. The two representations are related by
a unitary transformation which leaves the thermodynamic potential
unchanged and which therefore does not affect physical observables.
The change of basis is, however, reflected in the structure of the
diquark order parameter or, equivalently, in the expression for the
frequency selection rule.

Technically, temperature enters through the imposition of an
antiperiodic boundary condition on the fermion fields. There is a
certain degree of freedom in representing temperature terms in a
random matrix model. However, as soon as one imposes a given block
structure (e.g. diagonal or antisymmetric), one is actually selecting
a specific basis for the frequency states. As a result, attention must
be paid to the physical meaning of the basis states when solving for
the partition function and, accordingly, in the structure of the order
parameter. If a change of basis appears useful for mathematical
convenience, one must naturally keep track of the effects of such
transformation on the states and transform the order parameter
accordingly. From a general point of view, any transformation with the
block diagonal form of Eq.~(\ref{unitary-transformation}) with $U_m$
unitary leaves the thermodynamic potential invariant. There are thus
many equivalent Matsubara bases which can be chosen and
correspondingly many correct associated forms of $\omega_\pm$ and of
the diquark order parameter.  Among these forms, the antisymmetric one
appears to be particularly convenient since the corresponding diquark
projector is block diagonal in frequency space. The price to be paid
for working with a different form is merely the need to rely on the
knowledge of the physical meaning of the basis states and on which
frequency couplings are allowed. This will determine the form of the
corresponding diquark projector.

The situation is not very different for QCD with $N_c = 3$. If one
chooses a diagonal temperature term, one still needs to select
opposite frequencies for the fields involved in the diquark
condensate. In this case, however, the frequency selection mechanism
cannot be revealed by symmetry arguments alone since QCD with $N_c =
3$ does not satisfy an antiunitary symmetry.  As above, there are
many correct forms of $\omega_\pm$ which can be used together with the
corresponding prescriptions for the frequency couplings. In
Ref.~\cite{KleTouVer03}, the authors used the diagonal form of
Eq.~(\ref{T-omega3}) but did not consider diquark condensates. Thus,
there was no need for specifying a diquark projector, and the choice
of a diagonal form of $\omega_\pm$ was as correct as any other. On
the other hand, diquark condensation was investigated in
Ref.~\cite{VanJac00, VanJac01} using a diagonal form. There, it was
necessary to supplement the temperature dependence with an explicit
frequency selection rule.

In conclusion, the proper temperature dependence to be used in a given
random matrix model depends primarily on the form of the condensates that
may develop. Symmetry arguments are useful in identifying the
mechanism for selecting appropriate frequencies, but they are not always
sufficient.

\section*{Acknowledgments}

J. J. M. Verbaarschot is acknowledged for a valuable discussion.

\end{document}